\documentclass[
    ,final            % use final for the camera ready runs
  ]
  {aipproc}
  \usepackage{graphicx}

\layoutstyle{6x9}
\def\msun{M$_{\odot}$}
\def\mdot{$\dot M$}

\begin{document}

\title{Magnetic Accretion Onto White Dwarfs}

\author{Gaghik H. Tovmassian}{
  address={Observatorio Astr\'onomico Nacional, Instituto de Astronom\'{\i}a, UNAM, M\'exico\footnote{
P.O. Box 439027, San Diego, CA, 92143-9027, USA}}
}

% \author{<author2>}{
%  address={<common address for author2 and author3>}
% }
%
% \author{<author3>}{
%  address={<common address for author2 and author3>}
%  ,altaddress={<author1 address>} % additional visiting address
%}
%

\begin{abstract}
The influence of the magnetic field on  process of the accretion onto White Dwarfs in  Cataclysmic Variables (CVs)
is discussed. Except for the Polars or AM Her objects, the strength 
of magnetic field can not be measured directly in CVs by modern techniques. But there is  growing evidence that 
most of the types of Cataclysmic Variables classified on the basis 
of their observational characteristics are behaving in one or the other way under the influence of the 
magnetic field of the accreting White Dwarf, among other things. Here, we discuss the bulk of CVs that are traditionally
considered as  \emph{%
non magnetic} and review the properties that could be  best explained by the magnetic governed accretion process. 

\end{abstract}

\maketitle

%%%%%%%%%%%%%%%%%%%%%%%%%%%%%%%%%%%%%%%%%%%%
%% MAINMATTER
%%%%%%%%%%%%%%%%%%%%%%%%%%%%%%%%%%%%%%%%%%%%

\section{Introduction}

Cataclysmic Variables are at the  low mass end of objects known as Interactive Binaries,  
where the primary, more massive and accreting component  is a White Dwarf. The secondary 
is a pre-main sequence late type star that fills its Roche lobe and loses matter through the
inner Lagrangian point L$_1$ to its compact companion. The transferred matter forms an accretion disc
that usually is also the most luminous component of the binary system. Accretion is the cause of 
modulated brightness behavior with the Dwarf Novae (DN) class showing semi-periodic outbursts. According to the
traditional models of Cataclysmic Variables, the accretion disc forms as a 
result of exchange of the angular momentum between the elements or particles comprising the disc, which otherwise
move in Keplerian orbits in a ring with  its radius being uniquely determined by the angular momentum.
Thus, the ring (or torus) spreads out into a disk.  It is obvious then that 
between the inner edge of the disc and the surface of the accreting star, which rotates with a different velocity, 
the excess mechanical energy of disk's 
element must be dissipated and its excess angular momentum transferred away, before that element can 
be accreted onto the stellar surface. This region is called the boundary layer. The processes occurring
in the boundary layer and their observational fingerprints 
are not very well understood and remain a topic of controversy and debate in-spite of it
relatively simple general picture \cite{2001LNP...563..110S}. Since the discovery of the magnetic CVs \cite{1977ApJ...212L.125T},
it became obvious that at least some CVs possess a primary White Dwarf (WD) with a magnetic field strong enough
to disrupt formation of the accretion discs and channel the transfered material through the magnetic lines directly
onto the magnetic poles on the surface of the WD. It is also strong enough to overcome the spin-up torque of the accreting 
matter (see e.g., \cite{1991MNRAS.250..152K}) and synchronize rotation of the WD with orbital period.
Magnetic field strength of these can be measured directly from observations and subsequently they were called Polars. 
Later these were joined by Intermediate Polars (IP) which are notorious for showing spin period of asynchronously rotating
primary and often signs of presence of  accretion disc. The  magnetic field in IPs could not be measured directly, 
but it is unambiguously
established that the variety of periods observed there in X-rays and optical are result of spinning WD which
beams an intense X-ray emission modulated with P$_{\rm spin}$. It is also universally accepted that inner parts of the accretion
disk in IPs are truncated within corresponding Alfv\'en radius and the matter from there channeled onto the surface of the WD  
through the magnetic field lines, very much alike Polars. These two types of Cataclysmic Variables are 
defined as magnetic CVs and are the topic of another review talk at this conference \cite{2005AIPC}. Here, however,  I will 
show that many observed features in different sub-classes of CVs usually considered as non-magnetic can be generally 
explained in terms of truncated accretion discs as in IPs and that the magnetic governed accretion plays significant role 
across the entire family of Cataclysmic Variables.

\section{Why are White Dwarfs magnetic?}

It is natural to compare properties of isolated White Dwarfs and primaries of CVs in order to find out how binary evolution and
accretion processes influence their physics. And comparison of magnetic properties of these seemingly similar stars
reveals significant differences. 
Wickramasinghe and Ferrario \cite{2000PASP..112..873W} published a large study of magnetic White Dwarfs where they show  
that isolated 
WDs are remnants of Ap and Bp stars with fossil magnetic fields of order of ~0.1-1000 MG. They are thought to have 
significantly higher mean masses than their non-magnetic counterparts. They constitute about 5\% of WD population. 
In contrast White Dwarfs in interacting binaries do not reach so high magnetic fields, and their masses are not 
much different from overall distribution of masses in CVs. And even taking into account only CVs recognized as magnetic 
(Polars and Intermediate  Polars  $\approx10^5$ to $10^8$ Gauss) with significant and often measurable  strength of magnetic fields, 
the fraction of 
them easily reaches 25\% of the total. While the lack of very high magnetic field CVs is a matter of ongoing discussions and 
speculations, the disparity of numbers  can be attributed to  processes taking place in close binary systems, rather than 
selection effects.

More recently Aznar Cuadrado et al.\cite{2004A&A...423.1081A} discovered that probably up to 25\% of White Dwarfs posses 
low magnetic fields of a few kG. 
They were previously considered as non-magnetic. Tout et al. \cite{2004MNRAS.tmp..635T}  suggest that their magnetic field 
also has fossil origin of a cloud from which the stars emerged, and if so, all WDs born from stars over 2\msun might be a low 
field magnetic White Dwarfs. Regardless of the origins of magnetic field in WDs, it is important to stress for further 
consideration that the number of magnetic compact stars is much higher than previously thought.

\section{CV types that are suspected to be magnetic}
\subsubsection{SW Sex stars}

Similarly, the number of CVs in which magnetic driven accretion plays a significant role seems to be much higher than 
previously accepted.  
In most of the cases, it can not be measured directly, but there is observational evidence indicating influence 
of the magnetic field in the process of accretion on the WDs in a broad range of CV types. The most
obvious is the case of much debated SW Sex stars. They were distinguished \cite{1991AJ....102..272T}  for 
their peculiar emission and absorption line behavior. First thought to be eclipsing systems they were 
considered a rarity, but soon many other systems appeared showing one or other characteristics known as 
SW Sex phenomenon. According to Hellier \cite{2004RMxAC..20..148H},  more than 20 systems show SW Sex phenomenon.
Although he himself remains skeptical of the magnetic model for the explanation of SW Sex phenomenon, he admits that the 
evidence is mounting (see references therein). He assumes that periodic modulation of polarized emission is not enough evidence 
because the spin period of the WD does not come up in the photometric observations persistently, over a 
large time sets, as it easily happens in Intermediate Polars (IP).  But it also can be argued that our 
inability to observe spin/beat modulations  in SW Sex proves that there are many other CVs which do not 
show apparent IP  photometric characteristics, but do accrete under magnetic field influence. 

LS\,Peg  and V795\,Her, both members of SW\,Sex group are found to have circular polarization 
\cite{2001ApJ...548L..49R,2002pcvr.conf..533R}. 
This is direct evidence of magnetic 
field presence and its modulation with short periods most probably binds it with the spin period of WD. LS\,Peg shows 
circular polarization modulations with 
0.3\% amplitude and 29.6 min period. Simultaneously it shows emission line flaring with period corresponding to the 
beat period, if  29.6 min is considered  as a spin period of WD. Circular polarization  of 0.12\% peak-to-peak amplitude 
was also detected in V795\,Her with periods close to the optical quasi-periodic oscillations (QPO). 

But except detection of circular polarization there are many indirect indications of magnetic driven accretion onto white 
dwarf in SW\,Sex systems. The same kind of emission line flaring with short periods as in V795 Her 
is detected in a number of other SW\,Sex objects: BT\,Mon and DW\,UMa. They are  also typical to many Intermediate Polars 
(FO\,Aqr for example). A good example to demonstrate link between these seemingly different classes of CVs is V533\,Her 
\cite{2002MNRAS.337..209R}. It erupted as Nova in 1963. In 1979, Patterson \cite{1979ApJ...233L..13P} reported rapid 63.5 sec 
variability classifying it as a DQ Her system (magnetic). But later this period disappeared and the system emerged as a 
3.53 hour non-eclipsing SW\,Sex object showing among other SW\,Sex features flaring of emission lines.

Turning our attention to Nova remnants we find RR\,Cha, another Nova remnant that turned into an IP. 
Woudt \& Warner \cite{2002MNRAS.335...44W} discovered  a 1950\,sec stable period and positive and negative superhumps in the system. 
Meanwhile  Rodriguez-Gil and Potter \cite{2003MNRAS.342L...1R} observed variable circular polarization and noted  some distinct 
SW\,Sex features in its spectra.  Here we touch upon another phenomenon  (negative superhumps) that is common 
to number of systems. Patterson et al. \cite{2002PASP..114.1364P} in a study of yet another two SW\,Sex objects find QPOs 
with periods around 
1000\,sec  and negative superhumps. They believe the presence of negative superhumps can be best ascribed to 
the strong magnetism of the white dwarf. The warping of the disc is the most natural way of explanation of negative superhumps.
It is widely agreed that the cause of warping is a magnetic field. But there are different approaches to whether the source 
of the magnetic field is the primary \cite{1999ApJ...524.1030L} or the secondary \cite{2002MNRAS.335..247M}.

All these independent approaches to the SW\,Sex phenomenon show that the number of objects experiencing it 
are quite large and that the best explanation offered to explain variety of features is  magnetic accretion onto a 
white dwarf.

\subsubsection{VY Scl stars}

\begin{figure}[t]
\resizebox{7.5cm}{!}{
\begin{picture}(120,120)(0,15)
 \put (-45,160){\includegraphics[height=.24\textheight,angle=-90]{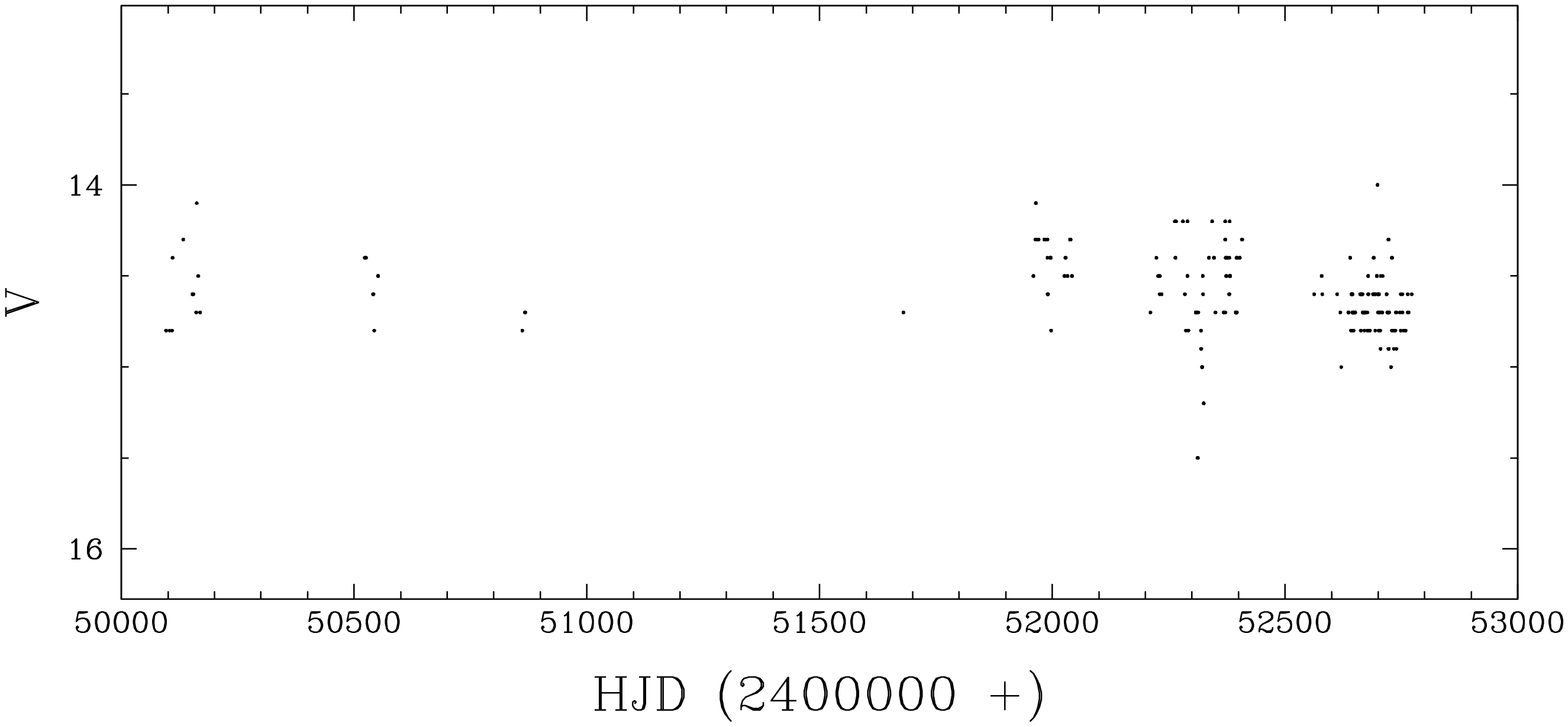}}
\put(100,70){\includegraphics[height=.15\textheight]{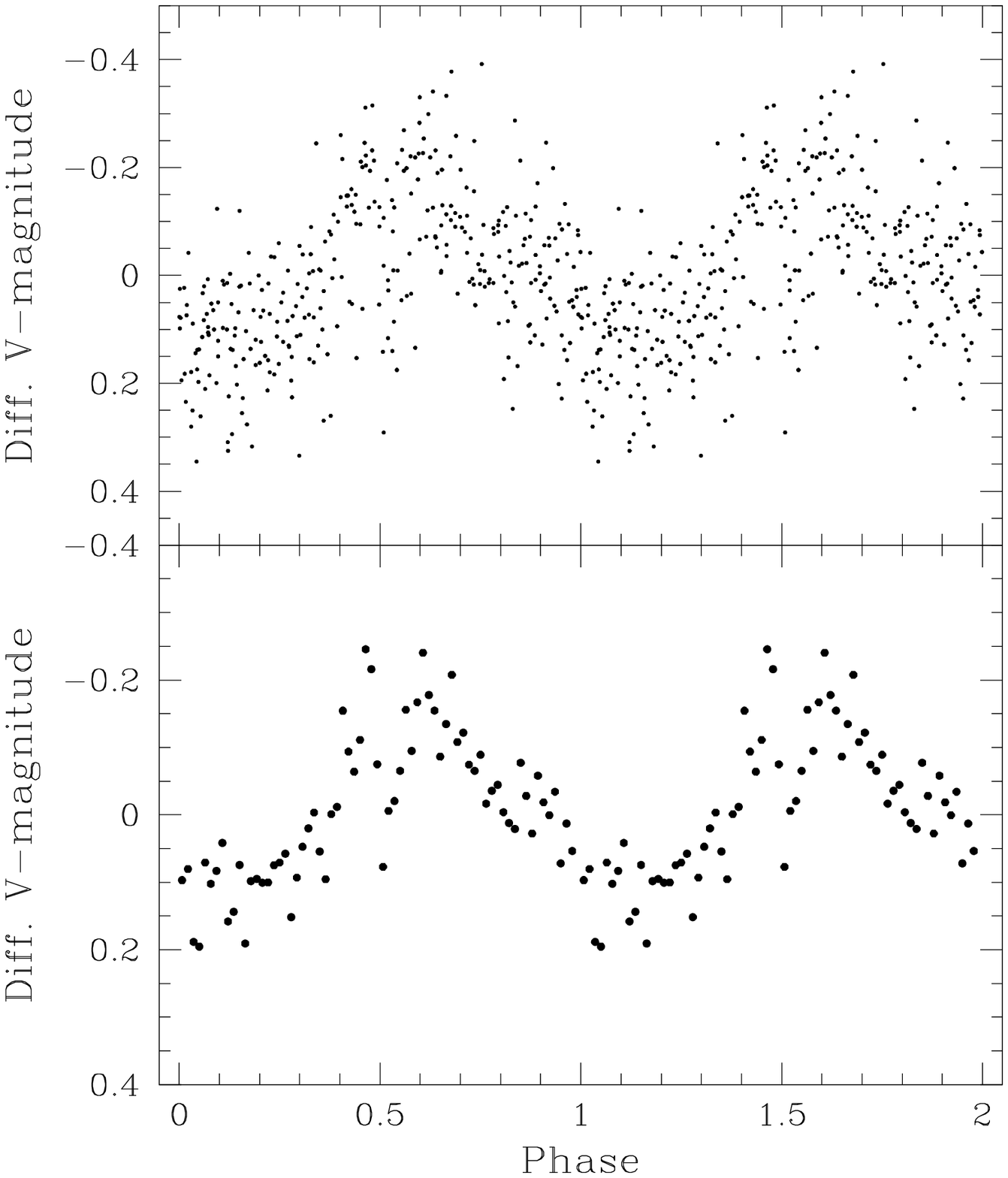}}
\put(-50,10){\includegraphics[height=.13\textheight]{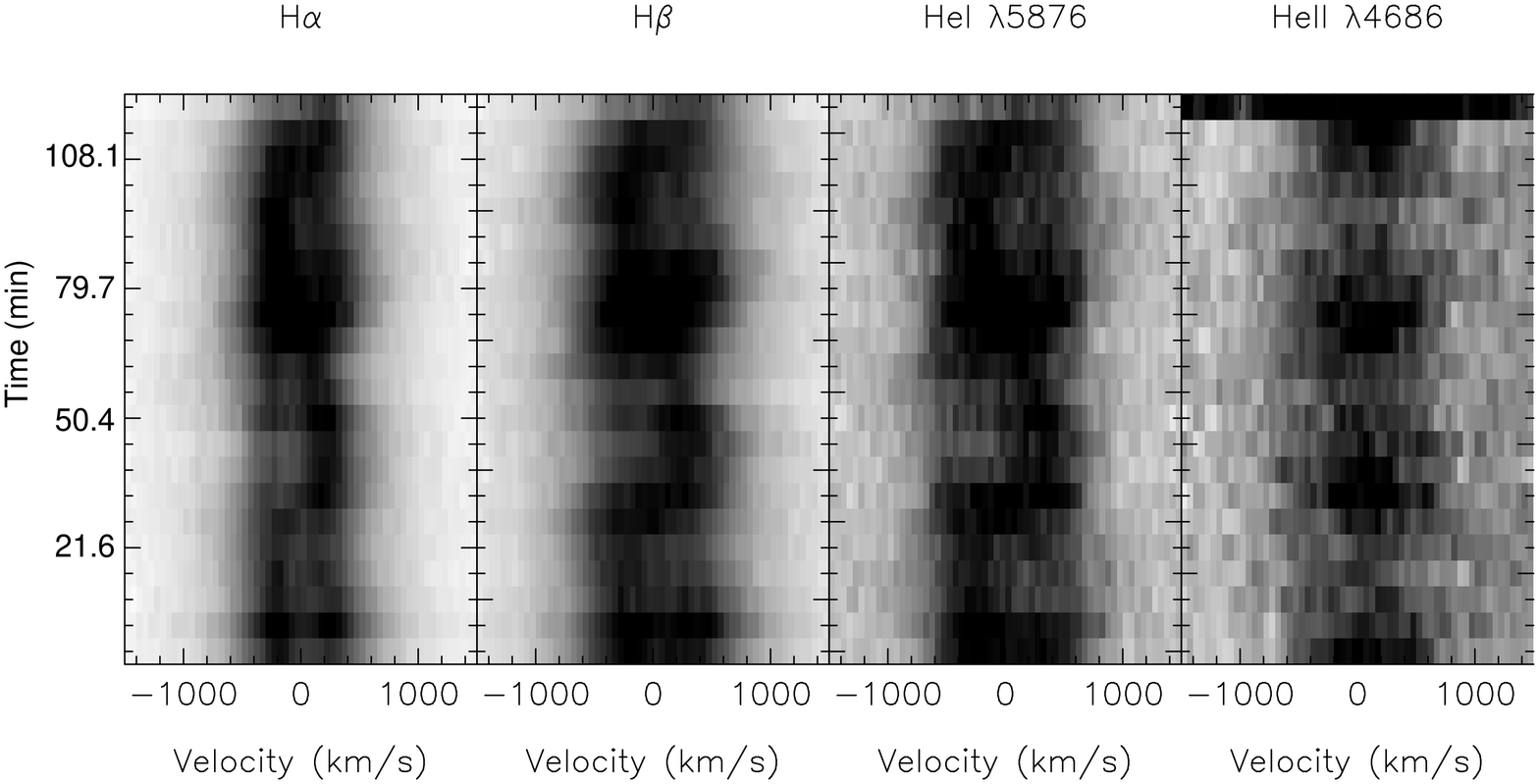}}
  \caption{DW Cnc exhibiting features proper to a)VY Scl objects ($\approx2$\,mag drop in brightness from quiescence levels); 
b) IPs (spin period of magnetic WD); c) SW SEX objects (emission line flaring). Adopted from  Rodriguez-Gil et al. (2004) }
\end{picture}}
\end{figure}

Among other features shown by SW\,Sex objects are so called VY\,Scl characteristics. Or rather SW\,Sex sometimes 
considered to be part of larger 
VY\,Scl type of CVs.  VY\,Scl  are another growing group of CVs that are increasingly associated with magnetic CVs. 
Here, it is first worth mentioning  DW\,Cnc. 
It was very recently studied by Rodriguez-Gil et al. \cite{2004MNRAS.349..367R}. They show that DW Cnc is a short period CV, 
P$_{\rm orb}=86$\,min, which also shows 
38.51 min photometric variability identified as a spin period of magnetic WD. Emission line flaring, another 
feature common to SW\,Sex objects, is present too.  Most interestingly, this short period CV  shows low 
states, down to 2 magnitude from its quiescence and no outbursts (see Fig. 1a-c borrowed from 
\cite{2004MNRAS.349..367R} exhibiting these features). The cyclical low states (anti dwarf nova behavior) 
is a main characteristic of VY\,Scl objects. They are believed to be concentrated in the 3-6 hours period range, 
same as SW\,Sex. A few years ago, we \cite{1999A&A...343..183G} demonstrated that the VY Scl star V751\,Cyg shows 
a transient soft X-ray emission. 
The X-ray emission appears when the system is in low state. The interpretation was that V751 Cyg behaves very much alike 
super-soft X-ray binaries, e.g.,  RX\,J0513.9-6951. Later, however, Hamuery \& Lasota \cite{2002A&A...394..231H} 
suggested that VY Scl objects 
contain magnetic WDs 
(with magnetic field of order of $5\times10^{30} {\rm G\,cm}^3$ corresponding to $0.4$ mG for 0.7\msun WD) based on their models 
and absence of outburst of VY Scl objects in low/intermediate states. 
This offers an alternative explanation of the soft X-ray emission in the low state. The spectrum of V751\,Cyg 
in a low state, obtained almost simultaneously with the X-ray observations, shows a spectrum similar to a mCV. 
It has highly variable continuum, He\,II  line becomes intense 
compared to the high state, lines are narrow, X-ray spectrum is soft. We  considered the magnetic/polar scenario 
based on spectral appearance  in the process of preparation of 1998 paper. The reason why we dismissed it, is 
still fundamental: how to switch off and on magnetic field or magnetic driven accretion between low and high 
luminosity states. 
Is it possible that with increased \mdot\ the magnetic field can not cope anymore with the amount of incoming 
matter and the accretion geometry changes? 
Hamuery \& Lasota are primarily concerned with conditions of accretion disc and argue that the truncation of 
the disk is a key to the VY Scl phenomenon, but they do not  reflect upon this question. But it needs to be 
answered. So far very little has been done to explore 
interaction and dependence of magnetic field to the mass transfer rate of accreted material. However there 
are hopeful signs that the problem 
can be tackled. Cumming \cite{2002MNRAS.333..589C} examined the problem to some depth. According to him compressional heating by 
accreting material can maintain interiors of WD in a liquid state. It allows to  a decrease  in the ohmic decay times 
to a few $10^9$ years in contrast to isolated WDs, where ohmic decay time 
is always longer than the cooling time. He shows that as a consequence of accretion significant changes in 
surface magnetic field can occur. 
He also demonstrates that the higher is the magnetic field of the system, the lower  the mass accretion rate. 
It is not immediately clear if the decrease 
of mass transfer/accretion rate in a VY\,Scl system provokes extension of magnetosphere and further truncation 
of the disc or just the decreased disc luminosity allow us to observe mCV features in the spectrum.  Or what  
a Polar would look like if you increase the accretion rate an order or two from usual 
$10^{11}$ \msun/year value.  Certainly this problem needs additional research.

VY\,Scl and SW\,Sex objects comprise considerable part of CVs at the upper edge of Period Gap. 
But the problem is not confined only to VY\,Scl or SW\,Sex objects. 
There were reports of apparently ordinary Dwarf Novae displaying features like those of 
VY\,Scl. Recently explored DW Cnc is just one such case. Another one was reported in 
\cite{1988AdSpR...8..329T,1986Ap.....24..131E}. SS\,Aur, a classical SS\,Cyg, system was caught in a low state for a 
short period of time. 
The object was down about a magnitude from its usual quiescence level. There were apparent changes in the spectrum
of the object: instead of a power law corresponding to the accretion disc spectrum, two blackbody curves  
corresponding to the stellar components of this binary system nicely fit the observed flux distribution.
The temperatures of components derived from this fitting were later confirmed from UV observations 
and parallax determinations by HST \cite{1999ApJ...515L..93H,2000AJ....120.2649H,2004AJ....128.1834S}.  
Most importantly it exhibited 
quasi-periodic photometric variations (Fig 2) with periods around 20 min. That is exactly what one would expect 
if the above described scenario is right: the  diminished disc luminosity either from decreased mass transfer
or truncation of the disc (or most probably from both at the same time) reveals periodic  light variations 
best explainable in terms of IP model. 

According to AAVSO light curves SS\,Aur was in a low state for very short time in contrast to usual VY\,Scl objects and
our observation of quasi-periodic variations at that moment was completely accidental. We may assume that other Dwarf Novae 
also experiment episodes of low state with short duration and they are  mostly unnoticed. It would be interesting to conduct a 
systematic search of such events and examine light curves for presence of periodic variations. Certainly it 
is necessary and would be easier to do that for members of VY\,Scl type of objects.

\begin{figure}[t]
\includegraphics[height=.35\textheight]{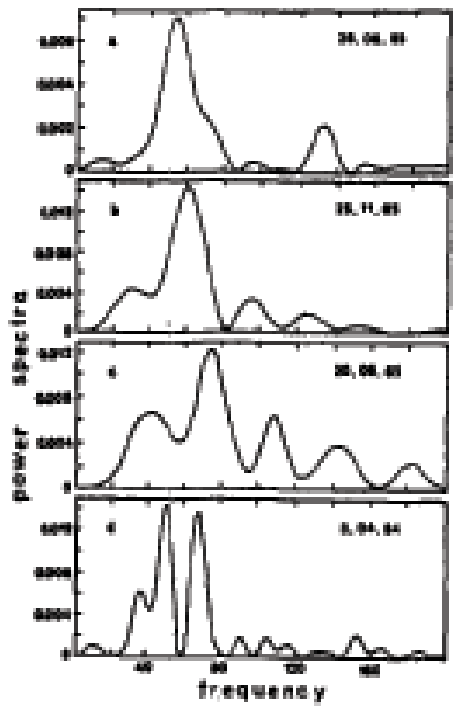}
\includegraphics[height=.34\textheight]{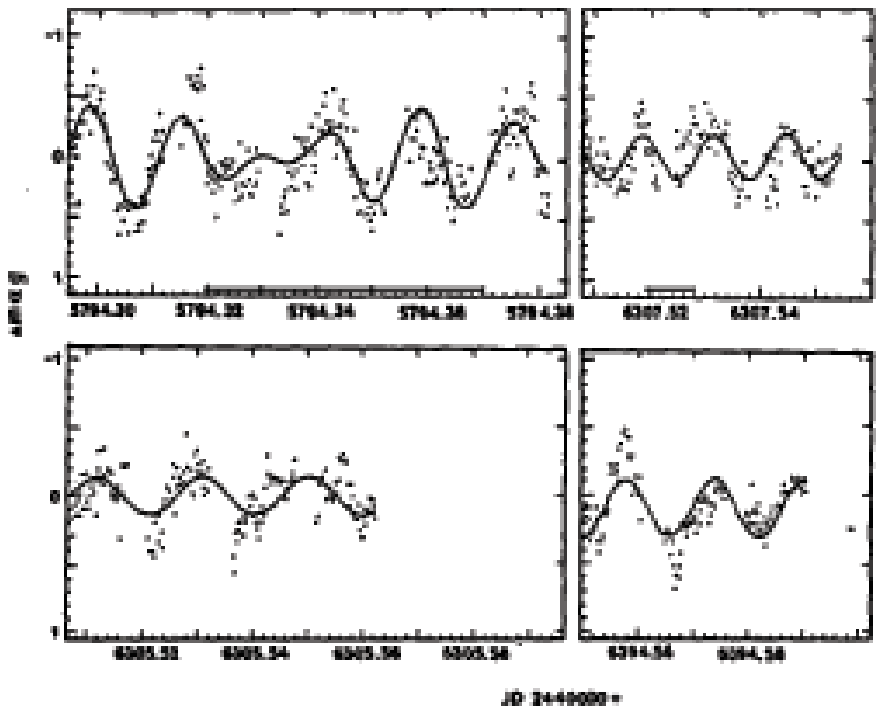}
  \caption{Quasiperiodic variations detected in the light curve of SS\,Aur during low luminosity states. 
On the left panel the power spectra of four different nights are presented, on the right corresponding light 
curves with $sin$ fits. Adopted from Tovmassian (1986).}
\end{figure}

\section{Other Dwarf Novae}

The DNe  constitute one of the most numerous group of objects among CVs.
The suggestion that many more  DNe  might experience short lasting low states (and their number is not limited to
a few known cases)  is completely speculative. However, another argument which favors the presence of 
magnetic driven accretion in DNe comes from observations of 
outbursts. Outbursts are the main feature that distinguish them from rest of CVs. Another well known fact is 
that during outbursts DNe show quasi-periodic oscillations  of three distinct types \cite{2004PASP.116.115}.
These oscillations have long been associated with boundary layer based on observation of eclipsing systems, 
but their nature was not clearly understood and described. The study of quasi-periodic oscillations is complicated
by the fact that they are not observed in every DNe, their magnitudes are small and highly variable and high time
resolution, high precision photometry is required.  One remarkable system that best suited for such study is VW Hydri.
Warner et al. \cite{2002MNRAS.335...84W,2003MNRAS.344.1193W}, Woudt \& Warner \cite{2002MNRAS.333...411} 
in series of papers present results of long term study of QPOs
mostly concentrated on this object, but not limited to it. They developed a Low Inertia Magnetic Accretor (LIMA) model
which allows to explain origin of QPOs and the existing  relation between  different types (different frequencies).
The essence of the model is that the rapidly rotating equatorial belt, formed as a result of accretion of matter 
through disc on a surface of WD, enhances magnetic field of the primary.
The magnetic field of a primary 
that is expected to be weaker than in regular Intermediate Polars, nevertheless reaches enough strength to channel 
accreting matter the way it does in IPs, but onto the equatorial belt instead of magnetic poles. 
The QPOs then arise due to a prograde traveling wave at the inner edge of the disc that reprocesses  high energy 
radiation from accreting zones close to the primary.  The frequency may be variable since the belt spins up during
high accretion phase and decelerates after. The details of this model and certain relation existing between
frequencies of QPOs common not only for CVs but also higher mass X-ray binaries are discussed in the Warner's \cite{2005AIPCW}
presentation included in this volume.

Interestingly, Huang et al. \cite{1996ApJ...458..355H} detected inverse P\,Cyg profiles during superoutburst of VW\,Hyi
and concluded that detached  disc  and structured gas flow is necessary for best-fitting model to describe their observation.
Subsequently in \cite{APJL1996.471.41} the same group demonstrate existence of equatorial belt around the WD after the outburst.
On the other hand, X-ray observations of high inclination 
system OY Car \cite{2003MNRAS.345.1009W} prove that X-rays come from an area much smaller than WD, probably  upper polar
region of the white dwarf, which testify that at least some DNe might have magnetic field strong enough to channel the accretion
to the magnetic pole.

Some theoretical aspects of how the magnetic field can be induced/enhanced by a shear and influence the processes 
in boundary layer were considered by Armitage \cite{2002MNRAS.330..895A}. Completely different 
approach  taken by Lasota \cite{2004RMxAC..20..124L} on the basis of Disc Instability Model (DIM) 
leads again to the idea that the internal 
parts of the accretion discs in most of DNe should be destroyed by the magnetic field and final stage of accretion occurs 
through magnetic lines. It is an extension of the idea first proposed for  VY\,Scl objects,  now  applied to the
OY\,Car, a classical Dwarf Nova for which the necessity  of truncated disc was raised earlier but was attributed to the
disc evaporation \cite{1994A&A...288..175M}. 

\section{ Conclusions}

There is a growing observational evidence that the number of magnetic WDs is larger than was thought. Observations 
of limited sample of isolated WDs show that as many as 25\% of WDs might have magnetic field strength of order of a few kG.
Regardless of  that fact, the number of systems considered as magnetic  in Interactive 
Binaries with WD as a  primary is unusually  high compared  to the distribution of isolated WDs. In addition to this,
there are numerous groups of CVs traditionally not considered as magnetic, which increasingly require the presence and influence
of the magnetic field on the accretion process in order to explain their observational characteristics. Probably the
Intermediate Polar scenario of accretion on WDs in CVs, where the inner disc is truncated and matter channeled 
to the primary along magnetic lines is  universal and accretion processes influence magnetic field strength in
accreting compact objects.  

Retter and Naylor 
\cite{2000MNRAS.319..510R} suggested that properties of CVs and thus their classification on both sides of 
Period Gap depends on their 
periods and mass transfer rate. Their scheme however does not include finer subdivision of CVs. If the 
hypothesis that the magnetic field plays important role in shaping properties of above-mentioned objects is correct, it could be stated 
that the classification of CVs is a function of their orbital period, mass transfer rate  and magnetic field strength.

% \begin{theacknowledgments}
% I would like to acknowledge continuous support of CONACyT through grants xxxxx and yyyy for study of magnetic CVs.
% \end{theacknowledgments}

%%%%%%%%%%%%%%%%%%%%%%%%%%%%%%%%%%%%%%%%%%%%%%%%
%% You may have to change the BibTeX style below, depending on your
%% setup or preferences.
%%
%% If the bibliography is produced without BibTeX comment out the
%% following lines and see the aipguide.pdf for further information.
%%
%% For The AIP proceedings layouts use either
%%%%%%%%%%%%%%%%%%%%%%%%%%%%%%%%%%%%%%%%%%%%

%\bibliographystyle{aipproc}   % if natbib is available
%\bibliographystyle{aipprocl} % if natbib is missing

%%%%%%%%%%%%%%%%%%%%%%%%%%%%%%%%%%%%%%%%%%%
%% You probably want to use your own bibtex database here
%%%%%%%%%%%%%%%%%%%%%%%%%%%%%%%%%%%%%%%%%%%

\end{document}